\documentclass{rmf-d}
\usepackage{nopageno,rmfbib,multicol,times,epsf,amsmath,amssymb,cite}
\usepackage[latin1]{inputenc}
\usepackage[]{caption2}
\usepackage[]{graphicx}
\usepackage{lipsum} 
\usepackage{array} 
\usepackage{nonfloat}
\newcommand\myfigure[1]{%
\medskip\noindent\begin{minipage}{\columnwidth}
\centering%
#1%
\end{minipage}\medskip}

\def\rmfcornisa{Research OR Education of Physics \hfill\rmf\ 
\hfill MES A\~NO}

\def\rmfcintilla{{\it Rev.\ Mex.\ Fis.\/} }
\clearpage \rmfcaptionstyle 
\begin{document}
\title{
Elastic meson form factors in a unified scheme
\vspace{-6pt}}
\author{Roger J. Hern\'andez-Pinto \footnote{Speaker} \ \footnote{roger@uas.edu.mx}}
\address{Facultad  de  Ciencias  F\'isico-Matem\'aticas,  Universidad  Aut\'onoma  de  Sinaloa,  Ciudad  Universitaria, CP 80000 Culiac\'an, M\'exico}
\author{M. Bedolla-Hern\'andez}
\address{Facultad de Ciencias en F\'isica y Matem\'aticas, Universidad Aut\'onoma de Chiapas,
Tuxtla Guti\'errez, 
Chiapas, M\'exico}
\author{L. X. Guti\'errez-Guerrero}
\address{CONACyT-Mesoamerican Centre for Theoretical Physics, 
\it Universidad Aut\'onoma de Chiapas, 
Tuxtla Guti\'errez, 
Chiapas, M\'exico}
\author{A. Bashir}
\address{Instituto de F\'isica y Matem\'aticas, Universidad Michoacana de San Nicol\'as de Hidalgo, Morelia, Michoac\'an,
M\'exico}


\maketitle
\recibido{day month year}{day month year
\vspace{-12pt}}
\begin{abstract}
\vspace{1em}
The extraction of elastic form factors for mesons in the context of the contact interaction model is revisited in this manuscript. The dressed masses of quarks and mesons are determined through the gap and Bethe-Salpeter equations. The generic elastic scattering process $M\gamma \to  M$ is studied for the meson $M$ formed of two differently flavored quarks. The charge radii of scalar, pseudoscalar, vector and axial-vector mesons are also extracted by virtue of explicit calculation of the meson elastic form factors.
\vspace{1em}
\end{abstract}
\keys{Quantum Chromodynamics, Mesons, Elastic Form Factors  \vspace{-4pt}}
\pacs{   \bf{\textit{12.38.-t, 12.38.Aw, 14.40.-n, 13.40.Gp, 24.85.+p}}    \vspace{-4pt}}
\begin{multicols}{2}

\section{Introduction}

The correct and precise description of hadron properties is a prohibitively difficult task. Perturbative quantum chromodynamics (QCD) helps us understand with great precision the Standard Model of particle physics in the regime of asymptotic freedom. However, hadrons are manifestations of other emergent properties of QCD, namely, the dynamical chiral symmetry breaking (DCSB) and confinement. A standard method to investigate the internal structure of hadrons is by the extraction of their form factors (FF). The prediction of perturbative QCD for meson FF, \cite{Lepage:1979zb}, shows that there exists $ Q_0>\Lambda_{\rm QCD}$, such that
\begin{equation}
\label{EqHardScattering}
Q^2 F_{0^-}(Q^2) \stackrel{Q^2 > Q_0^2}{\approx} 16 \pi \alpha_s(Q^2)  f_{0^-}^2 w_{0^-}^2(Q^2),
\end{equation}
where: $f_{0^-}$ is the meson's leptonic decay constant; $\alpha_s(Q^2)$ is the leading-order strong running-coupling
and
\begin{equation}
\label{wphi}
w_{0^-}(Q^2) = \frac{1}{3} \int_0^1 dx\, \frac{1}{x} \,\varphi_{0^-}(x;Q^2)\,,
\end{equation}
where $\varphi_{0^-}(x;Q^2)$ is the meson's dressed-valence-quark parton distribution amplitude. In this work, the elastic scattering process $M\gamma \to M$ between a meson $M$ and a photon $\gamma$ is analyzed, with the purpose of extracting information on the internal structure of mesons.

We start by providing the key ingredients for the calculation of the FF of mesons in the contact interaction (CI) model. Firstly, the dressed mass of the constituent quarks is calculated in this model. Then, the solution of the Bethe-Salpeter equation (BSE) to compute the meson masses is detailed. Finally, we extract the FF of mesons and 
determine their charge radii in this unified scheme. 

\section{Contact Interaction and the gap equation}
To study the internal structure of mesons in the $M\gamma \to M$ process, we first need to determine the dressed mass of the quarks which constitute the probed meson. In the CI model, it is calculated through solving the gap equation for the quark propagator, given by
\begin{align}
    &S(p,M_q)^{-1} = \imath \gamma\cdot p +m_q  \nonumber\\ 
    &+\int \frac{d^4q}{(2\pi)^4} g^2 D_{\mu\nu}(p-q)\frac{\lambda^a}{2}\gamma^{\mu}S(q,m)\frac{\lambda^a}{2} \Gamma_{\nu}(q,p) \, ,
    \label{eq:one}
\end{align}
where $S(p,m)=(p^2+m^2)^{-1}$ is the quark propagator, $M_q$ is the dressed quark mass, $m_q$ is its current value, $\gamma^{\mu}$ and $\lambda^a$ are the Dirac and Gell-Mann matrices respectively, $\Gamma_{\nu}$ is the quark-photon vertex and $D_{\mu\nu}$ is the gluon propagator. In perhaps the simplest truncation scheme, we consider $\Gamma_{\nu}\equiv \gamma_\nu$. To solve Eq.~(\ref{eq:one}) analytically, we use the CI. Hence the gluon propagator takes the form \cite{Gutierrez-Guerrero:2010waf},
\begin{align}
    g^2D_{\mu\nu} (p-q) = \delta_{\mu\nu}\frac{1}{m_G^2} \, ,
    \label{eq:two}
\end{align}
with $m_G$ being the gluon mass parameter. The solution of Eq.~(\ref{eq:one}) gives dressed masses of the quarks. They are found by solving the transcendental gap equation,
\begin{align}
    M_q =m_q+\frac{M_q}{3\pi^2m_G^2} \mathcal{C}^{iu}(M_q^2) \, ,
    \label{eq:three}
\end{align}
where $\mathcal{C}^{iu}(x)/x \equiv \Gamma(-1,x \, \tau_{\rm UV})-\Gamma(-1,x \, \tau_{\rm IR})$, with $\Gamma(a,b)$ the upper incomplete gamma function, and $\tau_{\rm IR}$ and $\tau_{\rm UV}$ are the parameters that regularize the integral \cite{Krein:1990sf, Roberts:2007ji}. 

\section{Bethe-Salpeter equation}
Meson masses are computed through the BSE. This equation, in the CI model, reads as \cite{Salpeter:1951sz},
\begin{align}
    \Gamma(k;P) =-\frac{4}{3}\frac{1}{m_G^2}\int\frac{d^4}{(2\pi)^4}\gamma_\mu \chi(q;P) \gamma^\mu
    \label{eq:four}
\end{align}
where $\chi(q;P)=S(q+P)\Gamma(q;P)S(q)$ and $\Gamma(q;P)$ is the Bethe-Salpeter amplitude (BSA), $k$ is the external relative momentum and $P$ the four momentum of the meson. Eq. (\ref{eq:one}) can be represented diagrammatically in Figure \ref{fig:one}.
\myfigure{%
\includegraphics[scale=0.55]{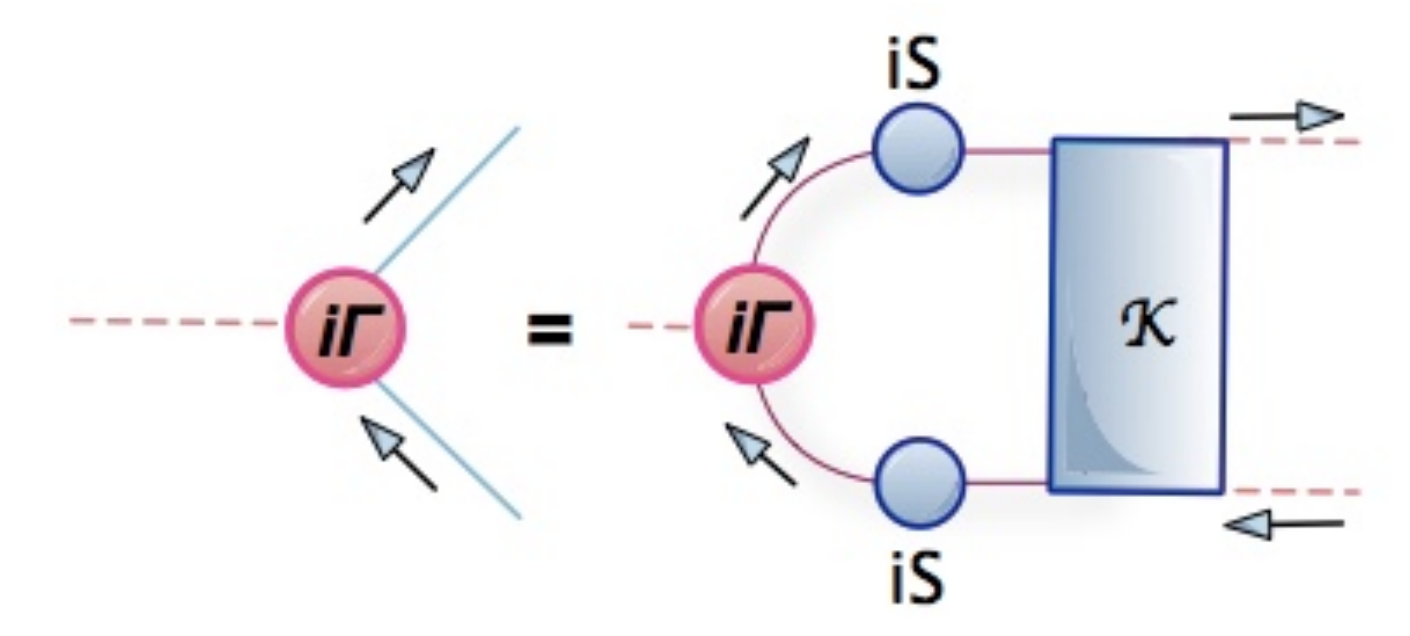}%
\figcaption{Diagrammatic representation of the BSE. Blue (solid) circles represent dressed quark propagators $S$, red (solid) circle is the meson BSA and the blue (solid) rectangle is the dressed-quark-antiquark scattering kernel $\mathcal{K}$.}%
\label{fig:one}%
}
We remark that the BSA are necessary for the calculation of the $M\gamma \to  M$ process in our approach. They encode the interaction properties between the meson and its constituent quarks. In the context of the CI, the BSA can be decomposed as,
\begin{align}
    \Gamma^H(P) = A^H(P) E^H(P) + B^H(P) F^H(P) \, ,
    \label{eq:five}
\end{align}
with $H={\rm S, PS, V, AV}$, representing the four kinds of mesons: Scalar (S), Pseudoscalar (PS), Vector (V) and Axial-Vector (AV). 

\vspace{3mm}
\begin{center}
\begin{tabular}{|ccc|}
\hline \hline
     BSA & $A(P)$ & $B(P)$  \\ \hline
     $\Gamma^{\rm S}$ & $\imath \,  I_D$ & -- \\ 
     $\Gamma^{\rm PS}$ & $\imath \,  \gamma_5$ & $\frac{1}{2M_R}\gamma_5 (\gamma\cdot P)$ \\
     $\Gamma^{\rm V}_\mu$ & $\gamma_\mu^{\rm T}$ & -- \\
     $\Gamma^{\rm AV}_\mu$ & $\gamma_5 \, \gamma_\mu^{\rm T}$ & -- \\ \hline \hline
\end{tabular}
\end{center}
Table 1: BSA for scalar, pseudoscalar, vector and axial-vector mesons. In here, $\gamma_\mu^{\rm T}$ satisfies $\gamma_\mu^{\rm T} P^\mu =0$.
\vspace{1mm}

The coefficients $A_H$ and $B_H$ are functions of Dirac matrices, $P$ and the reduced mass of the constituent quarks, $M_R=M_{q}M_{\bar{q}\prime}/(M_{q}+M_{\bar{q}\prime})$. The tensor structure used in this work is shown in Table 1.

In the case of pseudoscalar mesons the BSE can be written in terms of $E^{\rm PS}$ and $F^{\rm PS}$ as
\begin{eqnarray}
\left[ \begin{array}{c}
     E^{\rm PS}(P)   \\
     F^{\rm PS}(P)  
\end{array}
\right] = \frac{4\hat{\alpha}_{\rm IR}}{3\pi}
\left[ \begin{array}{cc}
     \mathcal{K}^{\rm PS}_{EE}   & \mathcal{K}^{\rm PS}_{EF} \\
     \mathcal{K}^{\rm PS}_{FE}   & \mathcal{K}^{\rm PS}_{FF}
\end{array}
\right]
\left[ \begin{array}{c}
     E^{\rm PS}(P)   \\
     F^{\rm PS}(P)  
\end{array}
\right]
\label{eq:six}
\end{eqnarray}
where $\hat{\alpha}_{\rm IR} = \alpha_{\rm IR}/m_g^2=1/(4\pi m_G^2)$, with $m_g=500$ MeV the mass of the gluon generated dynamically in QCD \cite{Gao:2017uox} and,
\begin{align}
    \mathcal{K}^{\rm PS}_{EE} &= \int_0^1 d\alpha \left\{  \mathcal{C}^{iu}(\omega_1) \right. \nonumber\\
    & \left. +(M_q M_{\bar{q}\prime} -\alpha(1-\alpha)P^2-\omega_1) \,  \overline{\mathcal{C}}_1^{iu} (\omega_1) \right\}, \\
    \mathcal{K}^{\rm PS}_{EF} &= \frac{P^2}{2M_R}\int_0^1 d\alpha ((1-\alpha)M_{\bar{q}\prime} +\alpha \, M_q) \overline{\mathcal{C}}_1^{iu} (\omega_1), \\
    \mathcal{K}^{\rm PS}_{FE} &= \frac{2M_R^2}{P^2}\mathcal{K}^{\rm PS}_{EF}, \\
    \mathcal{K}^{\rm PS}_{FF} &= \frac{1}{2} \int_0^1 d\alpha ((\alpha-1)M_{\bar{q}\prime}^2-M_q M_{\bar{q}\prime} -\alpha M_q^2) \nonumber \\
    &\times \overline{\mathcal{C}}_1^{iu} (\omega_1),
\end{align}
where $\omega_1 =M^2_{\bar{q}\prime} +\alpha M_q^2 +\alpha(1-\alpha)P^2 $ and $\overline{\mathcal{C}}_1^{iu} (z)=\mathcal{C}_1^{iu} (z)/z=\Gamma(0,z \, \tau_{\rm UV}) -\Gamma(0,z \, \tau_{\rm IR})$. Finally, the eigenvalue equation, Eq. (\ref{eq:six}), has a solution at $P^2=-m_H^2$.
In the case of vector, axial-vector and scalar mesons respectively, the BSE read as
\begin{align}
    \mathcal{K}_{\rm V}(P^2) & = \frac{2\hat{\alpha}_{\rm IR}}{3\pi} \int_0^1 d\alpha \mathcal{L}_{\rm V}(P^2) \overline{\mathcal{C}}_1^{iu} (\omega_1), \label{eq:eleven} \\
    \mathcal{K}_{\rm AV}(P^2) & = \frac{2\hat{\alpha}_{\rm IR}}{3\pi} \int_0^1 d\alpha (\mathcal{C}_1^{iu} (\omega_1)+\mathcal{L}_{\rm G}(P^2) \overline{\mathcal{C}}_1^{iu} (\omega_1)), \label{eq:twelve} \\
    \mathcal{K}_{\rm S}(P^2) & = \frac{4\hat{\alpha}_{\rm IR}}{3\pi} \int_0^1 d\alpha \left[\mathcal{L}_{\rm G}(P^2) \overline{\mathcal{C}}_1^{iu} (\omega_1) \right. \nonumber \\
    &-\left. \left( \mathcal{C}^{iu} (\omega_1) - \overline{\mathcal{C}}_1^{iu} (\omega_1) \right)
\right], \label{eq:thirteen}
\end{align}
where
\begin{align}
    \mathcal{L}_{\rm V}(P^2) &= M_{\bar{q}\prime} M_q-(1-\alpha) M^2_{\bar{q}\prime} \nonumber \\
    &- \alpha M^2_q -2\alpha(1-\alpha)P^2,\\
    \mathcal{L}_{\rm G}(P^2) &= M_q M_{\bar{q}\prime} +\alpha(1-\alpha)P^2.
    \label{eq:SPSAV}
\end{align}
The normalized solutions of Eqs. (\ref{eq:eleven})-(\ref{eq:thirteen}) provide masses of the vector ($m_{\rm V}$), axial-vector ($m_{\rm AV}$) and 
scalar ($m_{\rm S}$) mesons when,
\begin{align}
    1+(-1)^{i}\mathcal{K}_{j}(-m_j^2) = 0
    \label{eq:five}
\end{align}
with $i=\{0,1,0\}$ for $j = \{ \rm  S, V, AV \}$ respectively.

The procedure described above provides good phenomenological results in the case of vector mesons and flavour non-singlet pseudoscalar meson ground-states \cite{Chen:2012qr, Qin:2011xq,Maris:2006ea,Qin:2011dd, Cloet:2007pi}. For the parity partners, spin-orbit repulsion has to be taken into account \cite{Bermudez:2017bpx,Bashir:2011dp,Chang:2010hb,Chang:2011ei,Chang:2010jq}. In the CI, a phenomenological coupling $g_{SO}<1$ is introduced as a multiplicative common factor to the kernels. The numerical values for mesons with $J=0^+,1^+$ are \cite{Gutierrez-Guerrero:2021rsx},
\begin{align}
g_{SO}^{0^+} = 0.32 \, , \quad 
g_{SO}^{1^+} = 0.25 \,  .
\end{align}
To conclude this section, it is important to remark that the solutions of Eq. (\ref{eq:six}) and Eqs. (\ref{eq:eleven})-(\ref{eq:thirteen}) constrain the values of $E^H$ and $F^H$. These quantities in turn are crucial in subsequently computing the elastic FF of mesons.

\section{Elastic Form Factors of mesons}
Elastic meson FF  are calculated from the triangle diagram for $M\gamma \to M$, with a quark and an anti-quark circulating in the loop, see Figure \ref{fig:two}. We point out that in this article, the meson is considered to be made of two differently flavoured quarks.
\myfigure{%
\includegraphics[scale=0.25]{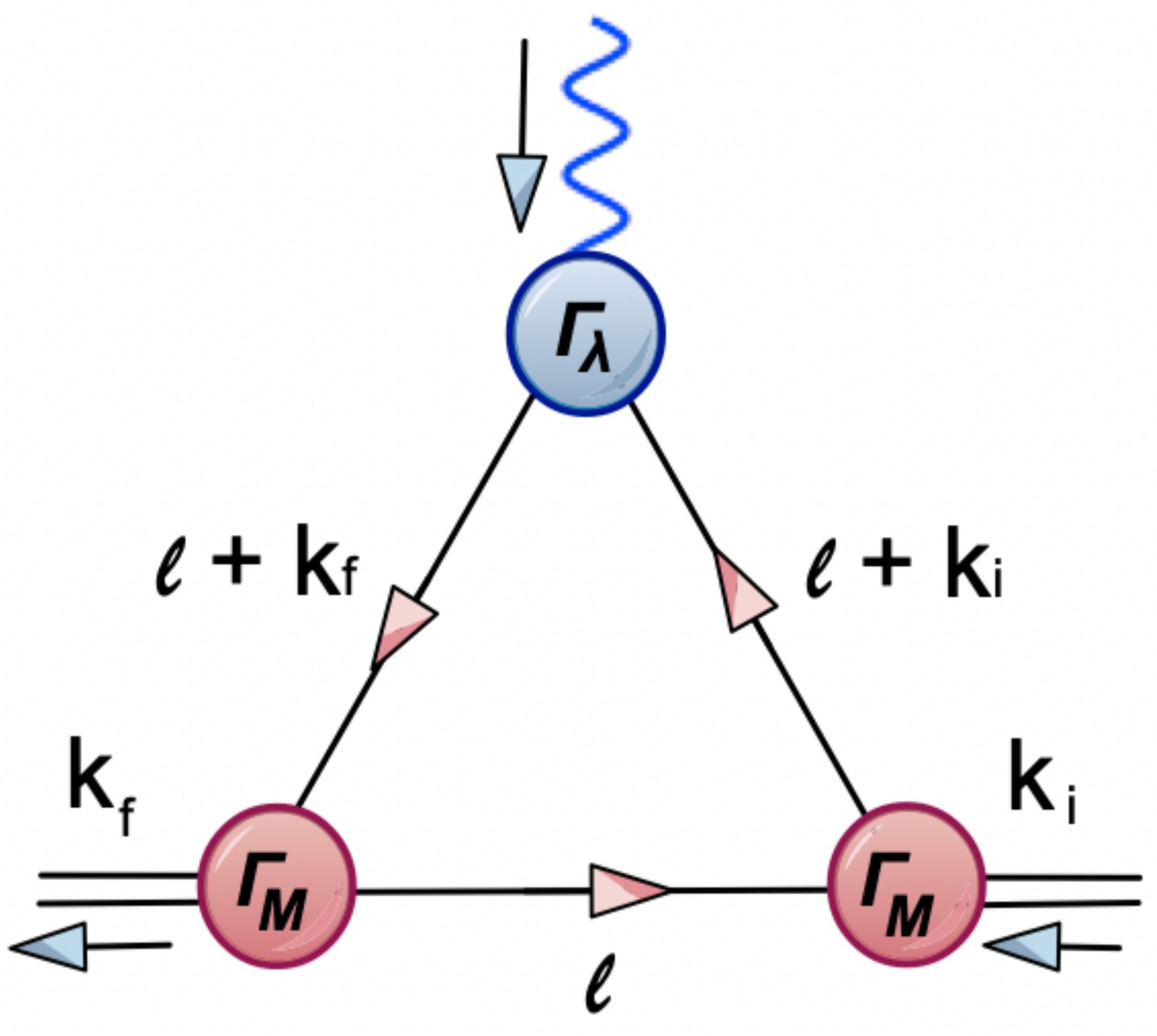}%
\figcaption{Feynman diagram for the $M\gamma M$ vertex which permits the extraction of elastic meson FF of all mesons. Labelling of internal momenta is included in the diagram.}%
\label{fig:two}%
}

Considering that the photon interacts with the quark while the anti-quark is a spectator, Feynman rules permit us to write this process as,
\begin{align}
\Lambda_{(\nu )\mu  (\alpha)} &= 2 \, N_c \,  {\rm Tr}\left[ \imath  \Gamma^{M}_{(\nu)}(-p_2) S_F(t+p_2,M_q)  \Gamma_{\mu} \right. \nonumber\\
&\times \left. S_F(t+p_1,M_q) \imath  \Gamma^{M}_{(\alpha)}(p_1) S_F(t,M_{\bar{q}\prime})\right]
\end{align}
where $\Gamma_{\mu} \equiv i P_T(Q^2) \gamma^{\mu}$ is the photon-quark dressed vertex, $P_T(Q^2)$ is the dressing function \cite{Maris:1999bh}, and the subscript between parentheses indicates that the meson may or may not have Lorentz indices, i.e. $\Lambda_\mu$ shall be understood for scalar and pseudoscalar mesons and $\Lambda_{\nu\mu\alpha}$ shall be used for vector and axial-vector mesons. The dressing function is given by, $P_T(Q^2) = (1+K_{\gamma}(Q^2))^{-1}$, where
\begin{align}
    K_{\gamma}(Q^2)=\frac{1}{3\pi^2 m_G^2}\int_0^1 d\alpha \, \alpha (1-\alpha) Q^2 \overline{\mathcal{C}}_1(\omega)
\end{align}
with $\omega=M_q^2+\alpha(1-\alpha)Q^2$.

In the case of scalar and pseudoscalar mesons, there is only one electromagnetic FF, $F_{\rm S, PS}^{\rm em}$. On the other hand, vector and axial-vector mesons have three form factors due to their tensor structure, $F_{\rm V, AV}^{\rm 1,2,3}$ \cite{Bhagwat:2006pu}. In terms of these FFs, the electric ($G_E$), magnetic ($G_M$) and quadrupole ($G_Q$) FFs are
\begin{align}
    G_E &= F^1 +\frac{2}{3}\eta G_Q \, , \quad G_M  = -F^2 \, ,\\
    G_Q &= F^1 +F^2 +(1+\eta)F^3 \, .
\end{align}
where $\eta \equiv Q^2/(4m_H^2)$ and $m_H$ is the meson mass.
After detailed analytical calculations, all FF can be written as,
\begin{align}\label{eqn:vecmes}
F^{j}_i(Q^2) &= \int_0^1 d\alpha \, d\beta \, \alpha \,  \nonumber \\ 
&\times \left(
 \mathcal{A}^j_i \, \overline{\mathcal{C}}_1(\omega_2) +(\mathcal{B}^{j}_i -\mathcal{A}^{j}_i \, \omega_2) \, \overline{\mathcal{C}}_2(\omega_2)
\right)
\end{align}
with $i= \{ \rm S, PS, V, AV \}$ and $j= \{ \rm em\}$ for the scalar and pseudoscalar mesons scenario and $j=\{ 1,2,3\}$ for the vector and axial-vector meson case. In addition, $\overline{\mathcal{C}}_2(z) = \left( \exp(-z\, \tau^2_{{\rm UV}}) - \exp(-z\, \tau_{\rm IR}) \right)/(2z)$ and $\omega_2=\alpha \, M_q^2+(1-\alpha)M_{q\prime}^2 -\alpha(1-\alpha)\, M^2_H +\alpha^2 \, \beta \,  (1-\beta) \, Q^2$. Naturally, the calculation of the FF also yields information of the magnetic and quadrupole moments for vector and axial-vector mesons, defined as
\begin{align}
    G^{i}_M(Q^2=0) \equiv \mu ^{i} \, ,\qquad G^{i}_Q(Q^2=0) \equiv \mathcal{Q}^{i} \, , 
\end{align}
with $i=\{\rm V,AV\}$, and these results can be also compared with experimental observation.
Note that in the case of mesons composed of differently flavored quarks, we define the meson FF, $\mathcal{F}_i^j$, as the sum of two contributions since the photon can interact either with the quark, $F_i^j$ or with the anti-quark, $\bar{F}_i^j$. Then \cite{Hutauruk:2016sug},
\begin{align}
    \mathcal{F}_{i}^j(Q^2) = e_q F_{i}^j(Q^2) + e_{\bar q} \bar{F}_{i}^j(Q^2)
\end{align}
with $e_q$ and $e_{\bar q}$ the quark and anti-quark charges respectively.
The procedure described up can readily be adapted for neutral meson with same flavored quarks.

Finally, once the analytic expression of the coefficients $\mathcal{A}$ and $\mathcal{B}$ are obtained, the comparison between the CI model and the experimental results can  also be carried out through the charge radii of different mesons, computed through the relation
\begin{align}
    r^2_{i} \equiv \Big\vert 6\frac{d}{dQ^2} F_i(Q^2)\Big\vert_{Q^2 =0} \, . \quad i=\{\rm S,PS,V,AV\}
\end{align}
This observable can constrain several models for mesons. An important feature of the FF is the dependence on $Q^2$. In Ref \cite{Gutierrez-Guerrero:2010waf}, it is pointed out that the behaviour of the pion electromegnetic FF at large $Q^2$ tends to a constant value in the CI.
For the rho meson, it turns negative for $Q^2\sim 6$ GeV. In this spirit, it would be interesting to delve into the behaviour of the FF of all kind of mesons in order to understand the predictions of the internal structure through the CI model.

\section{Conclusions}
In this document, the extraction of the elastic FF of scalar, pseudoscalar, vector and axial-vector mesons in the framework of the CI model has been presented in detail. 
The determination of the dressed quark and meson masses according to the gap and BSE helps constrain the parameters of the model. The study of the $M\gamma\to M$ process is presented in order to extract the elastic FF of mesons formed with two differently flavored quarks. Once the general expression is found, the charge radii of scalar, pseudoscalar, vector and axial-vector mesons can also be computed and compared with experimental results. Finally, interesting information such as the magnetic and quadrupole moments can also also be obtained in order to establish the extent of validity of the CI model more realistically.

\section*{Acknowledgments}
This research was supported by CONACyT through the Project No. A1- S-33202 (Ciencia Basica), Ciencia de Frontera 2021-2042 and Sistema Nacional de Investigadores. The work is also supported by PROFAPI 2022 Grant No. PRO\_A1\_024 (Universidad Aut\'onoma de Sinaloa), the COST Action CA16201 (PARTICLEFACE) and MCIN/AEI/10.13039/501100011033, Grant No. PID2020-114473GB-I00 and the CIC project 4.10 of UMSNH.

\end{multicols}
\end{document}